\documentclass[useAMS,usenatbib,oneside]{mn2e}
\usepackage{graphicx}
\usepackage{color}
\usepackage{amsmath,amssymb}
\def\xmm{XMM-{\it Newton}}
\def\sdss{{SDSS~J0959+1259}}

\title[Multiple AGN in the CG SDSS~J0959+1259]{Multiple AGN in the crowded field of the compact group  SDSS~J0959+1259}
\author
[A. De Rosa et al.]
{A. De Rosa$^{1}$\thanks{E-mail:alessandra.derosa@iaps.inaf.it}, 
S. Bianchi$^{2}$, 
T. Bogdanovi\'c$^{3}$, 
R. Decarli$^{4}$, 
R. Herrero-Illana$^{5}$,
B. Husemann$^{6}$, 
\newauthor
S. Komossa$^{7}$, 
E. Kun$^{8,9}$, 
N. Loiseau$^{10}$, 
Z. Paragi$^{11}$, 
M. Perez-Torres$^{5,12}$, 
E. Piconcelli$^{13}$, 
\newauthor
K. Schawinski$^{14}$,  
C. Vignali$^{15,16}$ \\
$^{1}$ INAF/Istituto di Astrofisica e Planetologie Spaziali. Via Fosso del Cavaliere - 00133 Roma - ITALY\\
$^{2}$ Dipartimento di Matematica e Fisica, Universit\`a degli Studi Roma Tre, via della Vasca Navale 84, 00146 Roma, Italy\\
$^{3}$ Center for Relativistic Astrophysics, Georgia Institute of Technology, Atlanta, GA 30332, USA\\
$^{4}$ Max-Planck Institut fuer Astronomie, Germany\\
$^{5}$ Instituto de Astrof\'{\i}sica de Andaluc\'{\i}a (IAA-CSIC), 18008 Granada, Spain\\
$^{6}$ European Southern Observatory Headquaters, Karl-Schwarzschild-Str. 2, 85748 Garching, Germany\\
$^{7}$ Max-Planck-Institut f{\"u}r Radioastronomie, Auf dem H{\"u}gel 69, 53121 Bonn, Germany\\
$^{8}$ Department of Experimental Physics, University of Szeged, D\'om t\'er 9, H-6720 Szeged, Hungary\\
$^{9}$ Department of Theoretical Physics, University of Szeged, Tisza Lajos krt 84-86, H-6720 Szeged, Hungary\\
$^{10}$ ESAC/XMM-Newton Science Operations Centre, Spain\\
$^{11}$ Joint Institute for VLBI ERIC, Postbus 2, NL-7900 AA Dwingeloo, The Netherlands\\
$^{12}$ Centro de Estudios de la F\'{\i}sica del Cosmos de Arag\'on (CEFCA), 44001 Teruel, Spain\\
$^{13}$ Osservatorio Astronomico di Roma (INAF), via Frascati 33, 00040 Monte Porzio Catone ( Roma), Italy \\
$^{14}$ Institute for Astronomy, Zurich, Switzerland\\
$^{15}$ Dipartimento di Fisica e Astronomia, Universit\`a degli studi di Bologna, Viale Berti Pichat
6/2, 40127, Bologna, Italy\\
$^{16}$ INAF-Osservatorio Astronomico di Bologna, Via Ranzani 1, 40127, Bologna, Italy
}

\begin{document}

\date{Accepted 2015 July 15.  Received 2015 July 8; in original form 2015 April 17}

\pagerange{\pageref{firstpage}--\pageref{lastpage}} \pubyear{2002}

\maketitle

\label{firstpage}

\begin{abstract}
We present a multi-wavelength study of a newly discovered compact group (CG), SDSS J0959+1259, based data from \xmm, SDSS and the Calar Alto optical imager BUSCA. With a maximum velocity offset of 500\,km s$^{-1}$, a mean redshift of 0.035,  and a mean spatial extension of 480 kpc, this CG is exceptional in having the highest concentration of nuclear activity in the local Universe,  established with a sensitivity limit L$_{X}>4\times $10$^{40}$ erg s$^{-1}$ in 2--10 keV band and R-band magnitude $M_R < -19$.
The group is composed of two type-2 Seyferts, one type-1 Seyfert, two LINERs and three star forming galaxies. Given the high X-ray luminosity of LINERs which reaches $\sim 10^{41}$ erg s$^{-1}$, it is likely that they are also accretion driven, bringing the number of active nuclei in this group to to 5 out of 8 (AGN fraction of 60\%). The distorted shape of one member of the CG suggests that strong interactions are  taking place among its galaxies through tidal forces. Therefore,  this system represents a case study for physical mechanisms that trigger nuclear activity and star formation in CGs.
\end{abstract}


%
%
\begin{table*}
 \caption{The sources of the CG in \sdss.}
\begin{tabular}{cccccccccccc}
\hline
 src & SDSS ID & RA  & DEC & z  &$\delta$v & ang. sep & Project dist. &  Lx & L([OIII])   & Type \\
&  & (1) & (2) & (3) & (4) & (5) &(6) & (7) & (8)  & (9)\\
\hline
 1 & J095906.68+130135.4 &  149.778   & 13.026 &    0.037 &    463 & 327  & 227 &   5 	&     0.07 &   Sy2 \\
 2 & J095903.28+130220.9 & 149.764    & 13.039  &  0.036  &  183  & 377  & 262 & 0.12 		&     0.003 &  LINER  \\
3 & J095908.95+130352.4 &   149.787   & 13.064  & 0.0339 &    317 &  318 &  221 & 0.1    &      0.0008 & LINER\\
4 & J095912.19+130410.5 &    149.801  & 13.070  & 0.0337 &    395 &   280  & 195 & $<$0.3	&    0.002 &   SFG \\
5 & J095914.76+125916.3 &   149.812   & 12.988  & 0.034 &    208 &  259 & 180 & 2.8 &   0.4 &    Sy2 \\
6 & J095859.91+130308.4  &   149.749   &  13.052  & 0.035 &    227 & 430  & 299 &   $<$0.2 &   0.008 &  SFG\\
7 & J095900.42+130241.6   &   149.752  & 13.045  & 0.035 &    20 & 418  & 290 &  $<$0.1 &     0.02 &  SFG \\
8 &  J095955.84+130237.7 & 149.983 & 13.043 & 0.035 & 30 & 389 & 270 & 15 & 0.03 & Sy1 \\
\hline
\end{tabular}

{{\footnotesize{$^{(1)(2)}$Right ascension and declination in degrees, position from SDSS; $^{(3)}$Redshift derived from our analysis; the uncertainty is of the order of 300 km/s. $^{(4)}$ Offset line-of-sight velocity in km/s measured with respect to the mean velocity 10,593 km\,s$^{-1}$; $^{(5)}$Angular separation in arcsec with respect to the center of the system (149.87$^\circ$;13.03$^\circ$). $^{(6)}$Projected distance in kpc with respect to the center assuming $H_0$ = 70km s$^{-1}$ Mpc$^{-1}$. $^{(7)}$Unabsorbed 2-10 keV luminosity in 10$^{42}$ erg s$^{-1}$ derived from our analysis. $^{(8)}$De-reddened (through Balmer decrement) [OIII] luminosity in 10$^{42}$ erg s$^{-1}$ derived from our analysis. $^{(9)}$ Object type (see Sect. \ref{crowded}).}}}
\label{names}
\end{table*}
%

\section{Introduction}

Compact groups (CGs) are systems composed of a small number of galaxies (three or more) in a compact configuration  with accordant redshifts, i.e., within 700km s$^{-1}$ from the group's mean velocity \citep{rose77,hickson94,hickson97}.
Some of the key open questions about CGs pertain to their relative importance in the universe and the relation between the global properties of these systems and the formation/evolution of their member galaxies (see the review by Hickson 1997). Because of their high galaxy densities, equivalent to those at the centers of rich clusters, and low velocity dispersions (about 200 km s$^{-1}$), CGs represent an environment where interactions, tidally triggered nuclear activity, and galaxy mergers are  expected to be more prevalent than in other environments. This is found in the famous cases of CGs Hickson 16 \citep[HCG~16;][]{ribeiro96}, Stephan's quintet \citep[HCG~92;][]{stephan77}, and Seyfert's sextet \citep[HCG~79;][]{seyfert51}.
However, the level of nuclear activity in CGs has not  been  understood yet. In a systematic study of 280 galaxies in 64 HCG \citet{martinez10} used optical emission-line ratios to classify the type of nuclear activity. They established that  28\% of sample galaxies  are AGN (37\%  being Seyfert-like and the remainder LINERs), 15\% are transition objects (TO), and 20\% are the star forming galaxies (SFG).  

Similarly, \citet{silverman14} used Chandra and XMM-Newton data to select 18 galaxy groups with mass M$_{group}\sim 10^{13}$ M$_{\odot}$ and redshift z $\sim$ 0.05.  The groups account for 2--30 galaxies each and were observed down to a sensitivity limit of L$_{X} >$ 10$^{40}$ erg s$^{-1}$ and B-band mag $<$18. They  found 16 AGN/LINERs distributed in 18 groups,  implying a mean number of AGNs/LINERs per group less than 1 (more precisely, 0.89$\pm$0.22).

The system discussed in this paper, associated with the \sdss\ at redshift z=0.035, has already been recognized as the only quintuplet detected in a sample of 1286 multiple AGN/LINER systems \citep{liu11}. 
The galaxies in this field that constitute our group have projected separations of $\lesssim$100h$^{-1}$ kpc and the line-of-sight velocity differences of $\lesssim$500 km s$^{-1}$ \citep{hickson97}. Within these criteria we find 7 spectroscopically identified sources which are listed in Table~\ref{names} (sources 1--7). We also find another source in this region (src8), which is close in redshift (z=0.035) but at larger projected separation, and include it in this study. 

The group in \sdss\ satisfies the criteria used by \citet{hickson97} to define of a CG: \\
(i) population: it is composed by more than 4 galaxies with absolute magnitudes within 3 mag of the brightest (Table \ref{busca});\\
 (ii) isolation: the radius of the smallest circle containing all the galaxies in the group (about 6.5$'$  in our CG, see Table \ref{names}) is at least three times smaller than the distance from the group center to the nearest unrelated galaxy with the absolute magnitude within 3 mag of the brightest member. Using NED database\footnote{https://ned.ipac.caltech.edu/} we find that the nearest such galaxy (with accordant redshift) is located at about 21' from the center of the CG;\\
(iii) compactness: the mean brightness surface of the group  in the R-band, calculated by distributing the flux of the member galaxies over the smallest circular area containing their geometric centers, is  $\sim$ 26 mag/arcsec$^{2}$ (29 mag/arcsec$^{2}$ including src8).

\begin{figure*}
\includegraphics[height=0.41\textwidth,width=6cm]{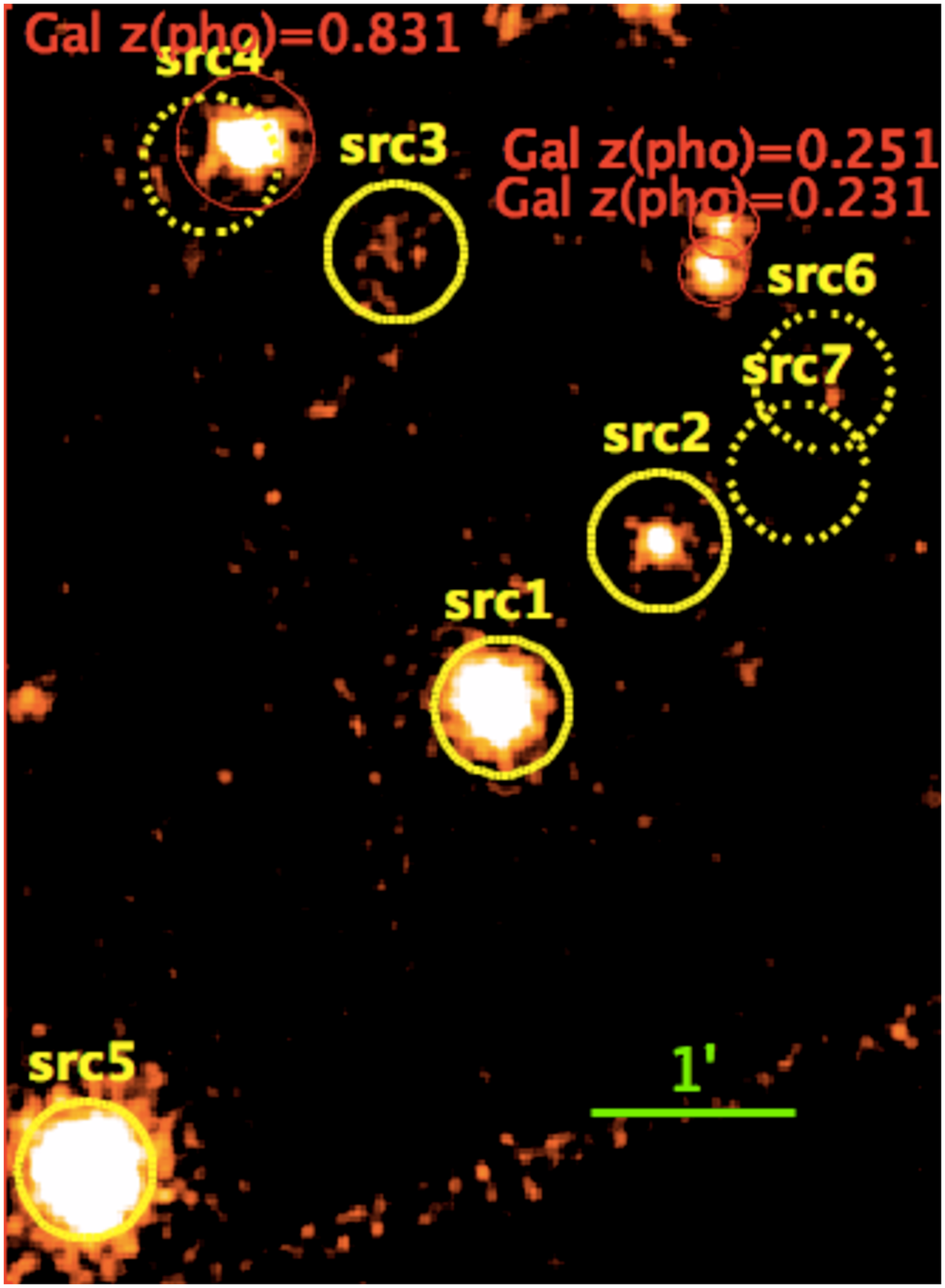}
\includegraphics[height=0.41\textwidth,width=6.cm]{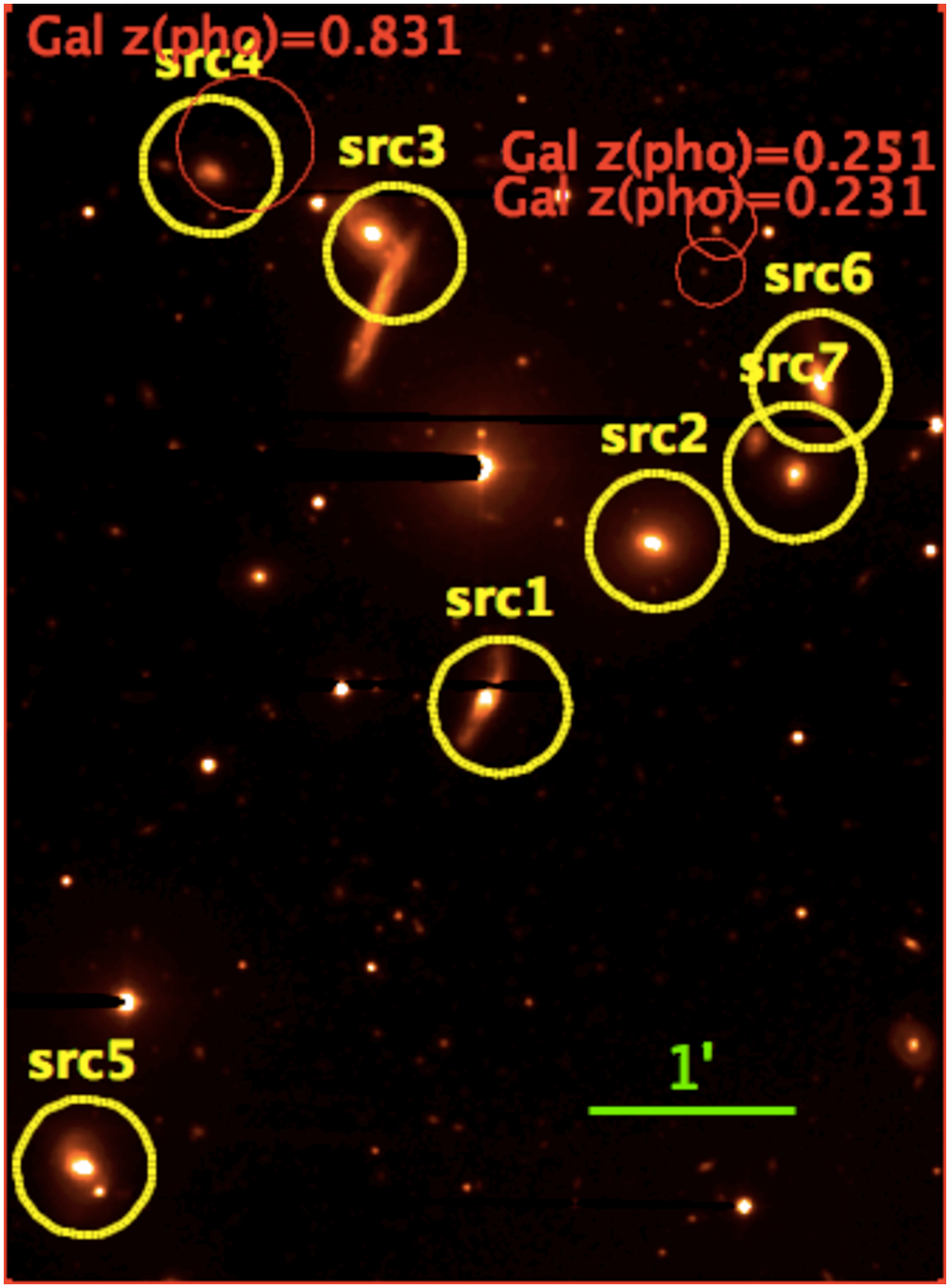}
\label{images2}
\caption{{\it Left:} XMM-EPIC smoothed mosaic image of the CG region (pn, MOS1 and MOS2 co-added). North is up and East to the left.  src 8 (NGC~3080) is off the plot, 10 arcmin to the East from src1. Yellow circles show the group sources listed in Tab.~\ref{names}. Undetected sources are shown with dashed line. Red circles are the sources in the field not included in our group, that lack a spectroscopic redshift in the SDSS (our estimated photometric redshifts based on BUSCA images suggest that they are background objects). {\it Right:} The  BUSCA R-band color image of the same region on the left. Labels are the same.
}
 \end{figure*}

The mean redshift of the CG is $\langle$z$\rangle$=0.0353 and the mean velocity is $\langle$v$\rangle$=10593 km s$^{-1}$. The center of this field has RA 09h 59m 28.97s and DEC 13$^\circ$ 01' 53.0'', and the average distance of sources from the geometric center of the group is 240~kpc. 
The X-rays (left) and optical (right) images of the CG are shown in Figure \ref{images2}.

In this paper we present a detailed optical and X-ray study of the CG in \sdss.
The observations and data analysis are presented in Sect. \ref{obs_sect} and Sect. \ref{crowded}, respectively, and results are discussed in Sect. \ref{disc_sect}.
Throughout this work, we assume H$_{0}$ = 70 km\,s$^{-1}$\,Mpc$^{-1}$, $\Omega_\Lambda$= 0.7, $\Omega_M$ = 0.3, and AB magnitudes. Errors and upper limits quoted in the paper correspond to the 90 per cent confidence level, unless noted otherwise.
%
%


\section{Observations and data reduction}
\label{obs_sect} 

\subsection{SDSS}
\label{sdsssec}

\begin{figure*}
\centering
\includegraphics[width=0.9\linewidth]{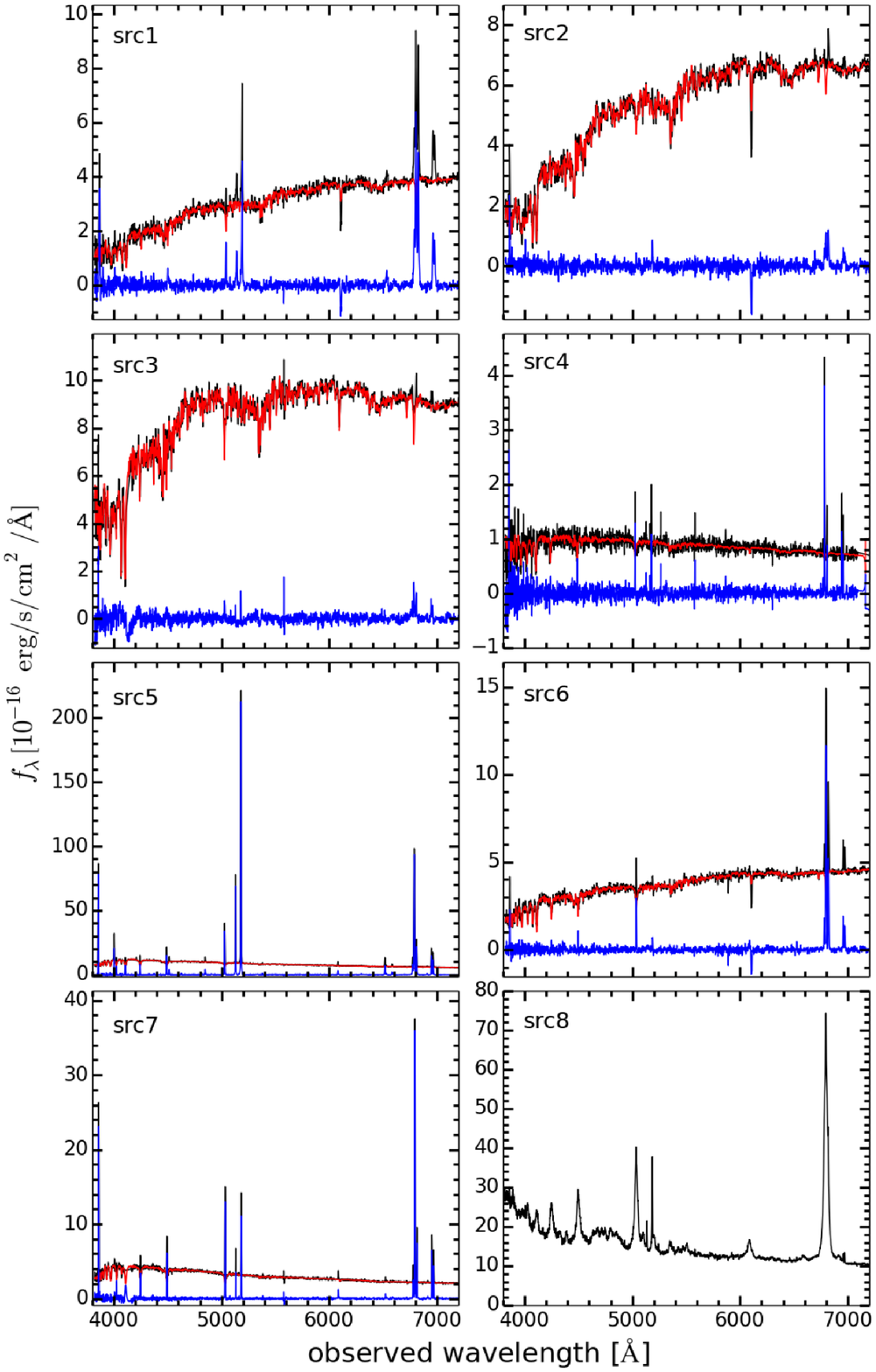}
  \caption{SDSS spectra of the sources in the field of \sdss. Composite spectrum (in black), model fit from the starlight subtraction (red), and starlight-subtracted spectrum (blue)}
  \label{sdss_spec}
\end{figure*}

All galaxies in our CG are targeted with SDSS under the legacy programme GALAXY \citep{strauss02}.
We retrieved the SDSS-III DR12 spectra for all them (see Table \ref{names}) from the survey webpage\footnote{http://skyserver.sdss3.org/dr12}.
The SDSS spectra are shown in Figure \ref{sdss_spec}.
The emission line flux of all the primary diagnostic lines (H$\beta$, [OIII] $\lambda5007$, [OI] $\lambda6300$,
H$\alpha$, [NII] $\lambda6583$ and [SII] $\lambda\lambda6713,6732$) were measured on top of the stellar continuum
using the package \textsc{PyParadise} (Husemann et al., in prep.) and listed in Table \ref{optical}. \textsc{PyParadise} models the stellar continuum as a superposition
of template stellar population spectra from the CB07 library \citep{bruzual03} after normalizing both the SDSS and the template spectra with a running mean over 100pix, interpolating regions with strong emission lines. A simple Gaussian kernel  is used to match the template spectra to the line-of-sight velocity distribution. The line fluxes are then inferred by fitting the Gaussian line profiles coupled in redshift and intrinsic rest-frame  velocity dispersion. Errors are obtained using a bootstrap approach where 100 realizations of the spectrum were generated based on the pixel errors with just 80\% of the template spectra and modelled again with the  same approach (at fixed stellar kinematics). 

\begin{figure*}
\begin{minipage}{0.8\linewidth}
\centering
\includegraphics[height=0.33\textwidth, width=13cm]{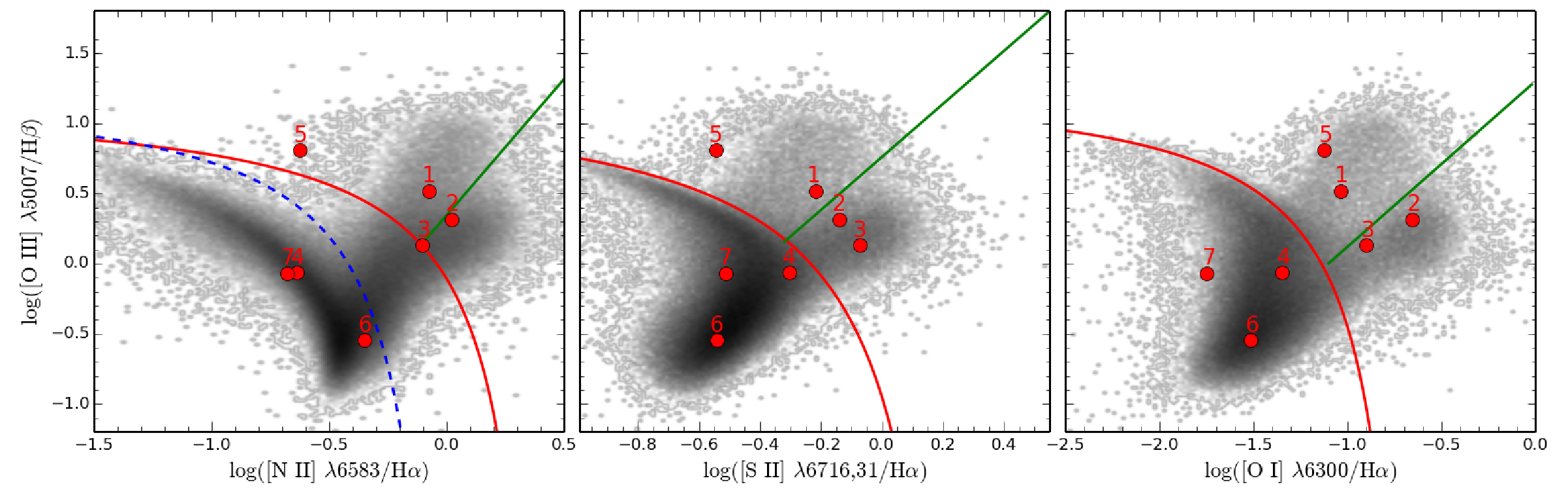}
\end{minipage}
\caption{Position of src1 to scr7 in the most common emission-line ratio diagnostic diagrams. Errors bars for these high S/N spectra are smaller than the symbol size. SDSS galaxies are shown in grey. Red solid line marks the maximum line starburst as inferred by \citet{kewley02}, the blue dashed line represents the empirical  SFG boundary proposed by \citet{kauffmann03}, and the green solid line is the  empirically proposed division between AGN and LINERs.}
\label{bpt}
\end{figure*}


\subsection{BUSCA}
\label{buscasect}

We observed SDSS J0959+1259 using the multi–band camera BUSCA on the 2.2\,m telescope in Calar Alto observatory. These observations span the whole optical window from U to I over a $12'\times12'$ field of view. The throughput curves of the dichroic mirrors, convolved with the detector efficiency, roughly correspond to those of the SDSS u, r, i+z, and Johnson B filters. Five 12 min-long frames were collected in each band, dithered by a few arcsec in order to clean our final images from cosmic rays and bad pixels. Data were processed with our own IRAF-based pipeline \texttt{redbusca}. The astrometric solution was computed via astrometry.net \citep{lang10}. The photometric calibration was obtained by comparing instrumental magnitudes of field stars with SDSS photometry, and assuming color corrections to account for differences in the filter throughput curves, as described in Mazzucchelli et al. (in prep.). 
The 10$\sigma$ detection limits for a point source are 25.76, 25.84, 25.56, 24.40 mag in U, B, R, and I, respectively. The seeing in R band was 1.0 arcsec. We created photometric catalogs using \texttt{SExtractor}
\citep{bertin96}. We base our source identifications on the R band image, which is more sensitive, and measure forced photometry on the other images based on the R-band input apertures.
We obtain the stellar mass M$_{\star}$ and star forming rate (SFR) estimates, listed in Table \ref{busca},  by fitting the available BUSCA+SDSS broad-band photometry using the SED fitting software MAGPHYS \citep{dacunha08}.
 BUSCA  observations are about a factor of nine deeper than SDSS, which has a limiting sensitivity of 22.0, 22.2, 22.2, 21.3 and 20.5 mag in u, g, r, i and z, respectively (the SDSS average seeing is about 1.43 arcsec).

\subsection{\xmm}
\label{xmm_analysis}
\xmm\ observed the field around \sdss\ on November 15, 2007 (ObsID: 0504100201) with a nominal exposure of 22 ks.
Both EPIC cameras (MOS and pn) were observing in full frame mode (thin filter). Data were reduced using SAS v13.5 with  standard settings and the most updated calibration files available at the time of the data reduction. 

The sources in the XMM field were detected using the EPIC source finding threads \texttt{edetect chain}, on 5 images in the $0.3-0.5$ keV, $0.5-1$ keV, $1-2$ keV, $2-4.5$ keV, $4.5-12$ keV energy bands with a detection threshold of 3$\sigma$. Sources src1, src2, src3, src5, and src8 are clearly detected in all bands. src8 and src5 have already been identified in X-rays as Seyfert-like AGN earlier \citep{lamassa09,baumgartner13}.  src6 and src7 are not detected (3$\sigma$ confidence level), while the X-ray emission visible in the \xmm\ composite image at a distance of $\sim$13 arcsec from src4 is likely due to a background quasar at z=0.831. 
All spectra were extracted from circular regions with 30$''$ radius which include  more than 80\% of the source counts at 1.5 keV in the EPIC cameras. The background spectra were extracted in the same CCD chip from circular regions free from contaminating sources and of the same size as the regions containing a source.

After screening selection and filtering for the flaring events, the net exposure for the EPIC cameras is 17.5 ks.
Each spectrum (and associated background) was rebinned in order to have at least 25 counts for each background-subtracted spectral channel and not to oversample the intrinsic energy resolution by a factor larger than 3.
The X-ray spectra of the sources detected with \xmm\ are shown in Figure \ref{spec}.
Spectral fits for pn and (co-added) MOS cameras were performed in the 0.3--10 keV energy band.

\begin{table*}
\caption{Optical emission lines. The fluxes are in 10$^{-16}$ erg cm$^{-2}$ s$^{-1}$ at the rest wavelength of the lines. See details in Section \ref{sdsssec}}
\begin{tabular}{ccccccccccccc}
\hline
 src & H$\beta$ &[O III] & [O III]  & [O I] & [N II] & H$\alpha$ &  [N II] & [S II] & [S II] & Type\\
  & 4861 \AA & 4960 \AA & 5007 \AA  &6300 \AA & 6548 \AA & 6563 \AA &6583 \AA & 6717 \AA & 6730 \AA & \\
  \hline
 1 &10.7 & 11.6 & 35.2 & 5.8& 17.5& 63.2  & 53.1 & 20.7 & 17.6 & Sy2\\
 2 & 3.4 & 2.3 &  7.1& 2.6&4.1 & 11.8 & 12.4 & 4.6 & 3.9 & LINER\\
 3 & 4.7 & 2.1 & 6.3 & 1.4 & 2.9 &  11.1 & 8.8 & 5.9 & 3.6 & LINER\\
 4 & 5.0 & 1.4&  4.3 &0.78  & 1.3& 17.6 & 4.0  & 5.4 & 3.4 & SFG\\
 5 & 162.9& 343.8 & 1041.8 & 41.7 & 43.5 & 555.5  &131.8 & 87.8 & 70.8 & Sy2\\
 6 & 15.9 & 1.5 &  4.5 &2.7 & 13.1& 88.9  & 39.6 & 14.3 & 11.3 & SFG\\
 7 & 51.9 &14.7 &  44.5  & 3.1 &12.0 & 173.4  & 36.4 & 31.7 & 21.6 & SFG\\
\hline
\end{tabular}
\label{optical}
\end{table*}
\begin{table*}
\caption{BUSCA data analysis results. See details in Section \ref{buscasect}.}
\begin{tabular}{ccccc}
\hline
 src & $\log M_{\star}$ & SFR & M$_{R}$  & M$_{I}$ \\
  & (1) & (2) & (3) & (4) \\
  \hline
 1 & 9.36 & 20 & -20.3 & -21.0 \\
 2 & 10.85 & $<$1 &  -21.1 & -21.8 \\
 3 & 10.52 & $<$1  & -20.7 & -21.3 \\
 4 & 9.30 & $<$1 &  -18.7 & -19.2  \\
 5 & 10.01 & 65 & -20.7 & -21.1 \\
 6 & 10.51 & 150 &  -20.4 & -21.1 \\
 7 & 9.29 & 5 &  -19.6  & -20.1 \\
\hline
\end{tabular}

{{\footnotesize{$^{(1)}$Stellar mass in  M$_\odot$; $^{(2)}$ Star forming rate in M$_\odot$/yr; $^{(3)}$ R-band absolute magnitude; $^{(4)}$ I-band absolute magnitude. }}}
\label{busca}
\end{table*}


\section{The crowded environment in the field of \sdss}
\label{crowded}

The CG \sdss\ is shown in Figure \ref{images2}, where the composite EPIC-pn and MOS12 (left panel) and the R-band BUSCA images  (right panel) are presented. src8 (NGC~3080), which is located 10 arcmin to the East from src1, is not shown. Yellow circles indicate CG sources, the dashed yellow circles mark the three objects not detected in X-rays, while the red circles show the background sources in the field not related to the group. 

The fluxes of the optical emission lines of the sources in the CG are listed in Table \ref{optical} and the most common emission-line diagnostic diagrams are shown in Figure \ref{bpt}. All three  diagrams lead to a consistent classification of objects into two Seyfert 2s, two LINERs and three SFGs. The optical spectrum of src8 shows broad Balmer emission lines (FWHM$\sim$ 750 km s$^{-1}$) and thus, points to a type-1 Seyfert. 

It is worth noting that our classification of src6 and src7 departs from that by \citet{liu11}, who report them as broad-line AGN. In order to examine the spectra for presence of the broad lines we fit the spectra with the narrow-line components and inspect the residuals. These show no evidence for broad lines. Moreover, we find that the narrow lines ratios point to a classification of src6 and src7 as SFGs (Fig. \ref{bpt}). It is possible that the mismatch is a consequence of a standard automatic procedure used by \citet{liu11} to classify a large number of sources in their sample.

\begin{figure*}
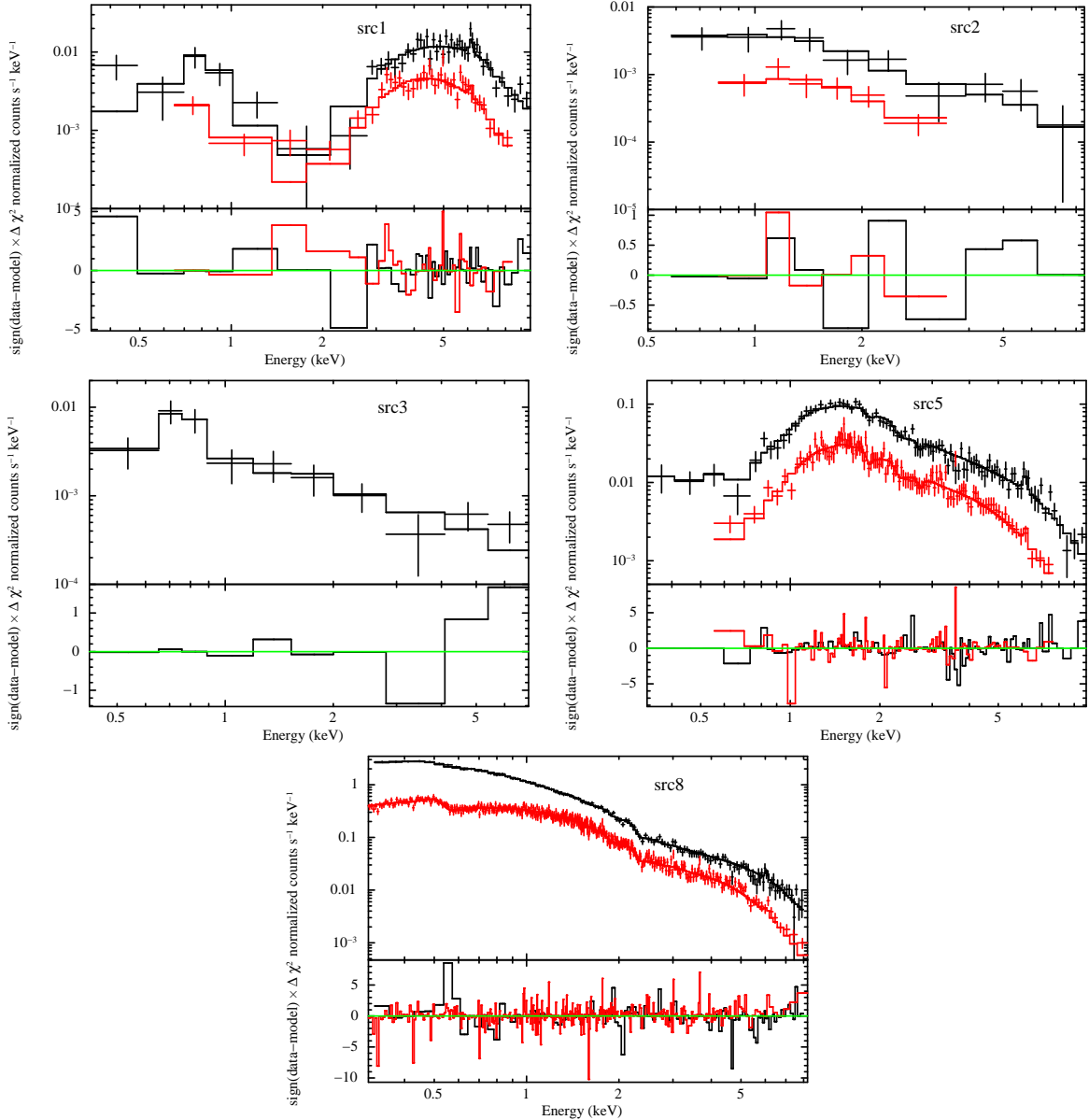

\centering
\includegraphics[width=0.3\linewidth, angle=-90]{src1_ldchisq.ps}
\includegraphics[width=0.3\linewidth, angle=-90]{src2_ldchisq.ps}
\includegraphics[width=0.3\linewidth, angle=-90]{src3_ldchisq.ps}
\includegraphics[width=0.3\linewidth, angle=-90]{src5_ldchisq.ps}
\includegraphics[width=0.3\linewidth, angle=-90]{src8_ldchisq.ps}
  \caption{XMM-pn (black) and MOS12 (red) co-added data of the sources detected in the region of \sdss. From top left src1, src2, src3, src5 and src8 is shown in the last panel. Bottom panels show residuals of the best fit models and our data set (see details in Sect. \ref{crowded}). No strong residuals are present across the whole energy band.}
  \label{spec}
\end{figure*}

The results of the X-ray spectral analysis are presented in Table \ref{fit} and best fit spectra are shown in Figure \ref{spec}.
The 0.3$-$10 keV spectra of  the two Seyfert 2s  (src1 and src5) are modelled with an absorbed power-law plus a thermal component with temperature of kT$\sim$0.1-0.6 keV (\texttt{mekal}  in Xspec) emerging below 2 keV. The cold absorption gas has column densities in the range $N_{H}\approx 1-20 \times 10^{22}$ cm$^{-2}$. In the Seyfert 1 source (src8) the broadband spectrum is well fitted with an absorption component with partial covering fraction f$_{c}$=0.41$\pm$0.07. The partial covering model is favoured with respect to a fully covering cold absorption with an F-test probability larger than 99.9 per cent.

A narrow Fe K$\alpha$ emission line is also measured in the three Seyferts, with an equivalent width of $\sim$100-130 eV. These are the brightest objects of the system in both the X-ray and optical wavebands.
The range of luminosity of the Seyferts is 10$^{42}$--10$^{43}$ erg s$^{-1}$, that is in a very good agreement with the values found in the studies of larger samples of local AGN \citep[e.g. CAIXA;][]{bianchi09}.

The two LINERs (src2 and src3) are modelled with a power-law continuum in the X-ray band. Both LINERs are unabsorbed and show no detectable Fe line,  although the S/N is poor above 5 keV due to their X-ray flux (less than 10 net counts in the pn). Their 2--10 keV luminosity is nevertheless high. At $\sim 10^{41}$ erg s$^{-1}$ it is above the mean value found in the systematic X-ray study of the largest sample of LINERs \citep{gonzalez09}. This study clearly shows that AGN sources have higher luminosities in $2-10$\,keV band ($\log{L_X}=40.22\pm1.24$) than non AGN-like sources ($\log{L_X}=39.33\pm1.16$). This is a strong indication that LINERs in this CG may be accretion driven.

None of the SFGs (src4, src6, src7) are detected by \xmm, placing a 3$\sigma$ upper limit on flux in 2-- 10\,keV of about 1.4$\times 10^{-14} {\rm erg\, cm^{-2} s^{-1}}$. This corresponds to a luminosity  of $\sim 4 \times10^{40} {\rm erg\, s^{-1}}$, assuming a photon index 1.7 and absorbing column denisity of 10$^{22}$ cm$^{-2}$.

\section{Discussion and conclusions}
\label{disc_sect}

The CG in \sdss\ represents one of the best examples of exceptionally strong nuclear activity in CGs in the nearby Universe. The combined optical and X-ray analysis clearly shows that this  system is formed by two type 2 Compton-thin Seyferts, one type-1 Seyfert, two LINERs, and three SFGs. 

In terms of spectral components and shape, the X-ray behaviour of LINERs (src2 and src3) is in very good agreement with the global properties of the sample of LINERs investigated in X-rays by \citet{gonzalez09}. However, in terms of  2--10 keV luminosity, both src2 and src3 show high  values ($\sim10^{41}$erg s$^{-1}$) that is within the average value found for the AGN candidates.

We conclude that our LINERs are likely accretion driven, which increase the fraction of AGNs in this CG to from 40 to 60\% (from 3 to 5 out of 8). The only other example of an AGN rich group like this is the well known HCG~16 \citep{ribeiro96,turner01}, with an AGN fraction of 75\% (three out of four galaxies).

Note that fiber collision bias in SDSS survey is an important effect \citep{li06}, 
in fact there is a limit of 55$''$ to how close two fibers can be on the same tile.
Consequently $\sim$ 10\% of targeted galaxies from a photometric catalog cannot be assigned fibers and obtain measured spectroscopic redshifts. However, all brightest galaxies of the CG discussed here were spectroscopically observed by SDSS, implying that we cannot be missing AGN because of fiber collisions. On the other hand, other groups with similar AGN fraction may remain undetected due to this well-known effect. In the present work we are not aiming at a statistical study of the incidence of CG with high AGN fractions, but rather reporting the detection of such a unique group.

We compared our result with a systematic analysis of a sample of CGs with mass M$_{group}\sim 10^{13}$M$_\odot$, investigated in X-rays with a limiting luminosity L$_{X} >$ 10$^{40}$ erg s$^{-1}$ and B-band mag $<$18 \citep{silverman14}. The number of AGN in our group is significantly higher than the mean number of 0.89$\pm$0.22 AGN/LINERs per group found by \citet{silverman14},  with a similar L$_{X}$ and M$_R$ selection. The AGN fraction found in their analysis was 36$^{+14}_{-11}$\% and 13$^{+9}_{-4}$\% for galaxies classified as central or satellite, respectively.

In order to find how common is the level of activity found in the GC analysed here we also compare it to several other environments at low and high redshift. 
For example, rescaling our results for their adopted X-ray luminosity and R-band magnitude limits (L$_{X}>$ 10$^{42}$ erg s$^{-1}$ and M$_{R}<$-20), we find that the our AGN fraction is 40\%, i.e.  higher than the estimates of the AGN fraction of galaxies in the field  \citep[1.19$\pm$0.11\%;][]{haggard10}.
Our AGN fraction is also higher than that measured in isolated galaxies \citep[7-20\%; ][]{sabater08}.
Using similar magnitude selection, and rescaling our results for their adopted X-ray luminosity limits, 
the AGN  fraction in \sdss, is 60\%, i.e.  higher than found in clusters \citep[2-5\%, with M$_R<$-20 and L$_{X}>$10$^{41}$ erg s$^{-1}$;][]{martini06}.
In high-z  galaxies (0.25 $<$ z $<$ 1.05) with mass M$_{\star}>$2.5$\times 10^{10}$\,M$_{\odot}$, the AGN fraction with a X-ray luminosity limit L$_{X}>$ 10$^{42}$ erg s$^{-1}$ is of about 10\% \citep{silverman11}. These galaxies are commonly found in kinematic pairs characterized by physical separations less than 75\,kpc and a line-of-sight velocity difference less than 500\,km\,s$^{-1}$. 
In \sdss, the  fraction of AGN with L$_{X}$ above 10$^{42}$ erg s$^{-1}$ is 40\%, significantly higher with respect to the sample of high-z galaxies.
 
Using SFR \textit{vs} stellar mass relation at z=0, we find the values of SFR for src6 and src7 (see Table \ref{busca}) higher than measured in the local Universe \citep{elbaz07}. This indicates an enhanced star formation, which is especially evident in src6. In addition, the fraction of the SFG galaxies in this CG is 38\% (3 out of 8), while in HCGs the percentage is around 20\% \citep{martinez10}.

The values of the specific star formation rate  (which is a measure of the relative growth-rate of the galaxy) obtained with BUSCA data for the Seyfert 2s, is about 2.5 Gyr$^{-1}$ (see Table \ref{busca}), in good agreement with  values measured in larger samples of AGN in the local Universe \citep{rovilos12, elbaz07}.
The detection of a thermal component in the Seyfert 2s is highly significant, at $>$99.9\% confidence (see Table \ref{fit}), and it contributes about 1--2\% to the total luminosity in 0.5--2 keV ($L_{th}\sim$2--5 $\times$ 10$^{40}$erg s$^{-1}$).  This component is likely due to the emission from the Narrow Line Regions \citep{bianchi07}, however, we cannot exclude a possibility that it is due to a nuclear SF component. Using the relation from \citet{ranalli03}, we  obtain an upper limit for the nuclear SFR of about 20 and 10M$_\odot$ yr$^{-1}$ for src1 and src5, respectively. 

The Arecibo HI survey ALFALFA \citep{giovanelli05} shows a very large amount of HI gas in this group (2$\times10^{10}$M$_\odot$).
It is unclear whether its origin is intergalactic, as found in other CGs \citep{koribalski03}, or  intragalactic, but it apparently could provide enough material to fuel the AGN and SFGs. In this scenario, the Broad Line Region detected in src8 can in principle be formed by gas enrichment. The ALFALFA survey has recently been used to show that HI has disturbed morphologies (tails and bridges) in post-merger galaxies while it exhibits an abundance similar to isolated galaxies \citep{ellison15}. 
The distorted shape of the edge-on disc galaxy (src1 in  Figure \ref{images2}) already suggests that strong interactions are occurring among the galaxies in the CG mediated by the tidal forces or ram pressure of the intra-group medium. Furthermore, src5  exhibits an extraordinary low  [NII]/H$\alpha$ ratio (the error on this ratio for type 2 AGN is much lower than the spectral offset found, see Fig. \ref{bpt}), but is clearly a Compton-thin AGN. 
This is very rare in the local Universe \citep{groves06}, and it could be due to a recent galaxy interaction or due to accretion of low-metallicity gas from the intragroup environment.

The high incidence of AGN in this CG point to favourable conditions for inducing the black hole growth. 
However, the accretion rate in M$_{\odot}$/yr estimated from accretion luminosity (L$_{X}$) is relatively low, so the activity cannot be simply explained by the presence of gas.  The enhanced nuclear activity, the presence of SFGs and the proximity of the CG allow detailed, spatially resolved mapping of the distribution and kinematics of the stellar and gaseous components with the Multi Unit Spectroscopic Explorer (MUSE, with a proposal by our team already approved). This makes the system a good case study for physical mechanisms that trigger AGN and star formation in CGs.

\begin{table*}
 \caption{X-ray spectral analysis. Combined EPIC-pn and MOS12 data. See details in Sect. \ref{xmm_analysis}.}
 \begin{tabular}{ccccccccc}
 \hline
src &  counts & $N_{H}$ & $\Gamma$ & $kT$ & EW (Fe K$\alpha$) & $L_{\rm soft}$ & $L_{\rm hard}$ & F$_{2-10 keV}^{obs}$\\
 & (1) & (2) & (3) & (4) & (5) & (6) & (7) & (8)\\
\hline
1 & 5.8$\pm$0.2 & 13$^{+2}_{-2}$ & 1.5$\pm$0.3 &  0.6$^{+0.1}_{-0.2}$ &100$\pm$60 & 2.7$\pm$0.1 (2\%) & 5.3$\pm$0.1 & 0.89$\pm$0.01\\
2 & 0.9$\pm$0.1 & $<$0.2 & 1.3$^{+0.6}_{-0.4}$ & - & - & 0.04$\pm$0.01 & 0.12$\pm$0.05 & 0.04$\pm$0.02 \\
3 & 0.77$\pm$0.09 & 1.0$^{+1.0}_{-0.6}$ & 1.7$^\star$ & 0.4$^{+0.2}_{-0.1}$ & - & 0.08$\pm$0.01 (25\%) & 0.11$\pm$0.02 & 0.04$\pm$0.01\\
5 & 20.1$\pm$0.4 & 0.77$^{+0.06}_{-0.06}$  & 1.91$\pm$0.08 & 0.11$^{+0.07}_{-0.02}$ & 130$\pm$80 & 2.2$\pm$0.1 (1\%)  & 2.8$\pm$0.1  & 0.94$\pm$0.03\\
8 & 220$\pm$1&  $^{\dagger}$14$^{+5}_{-4}$& 2.41 $\pm$0.02 & - &  73$\pm$40 & 22.6$\pm$0.1 & 14.7$\pm$0.2 & 3.7$\pm$0.1\\
\hline
\end{tabular}
\\
\footnotesize{{$^{(1)}$ Total 0.3-10 keV pn counts in 10$^{-2}$s$^{-1}$  ;$^{(2)}$Absorption column density in 10$^{22}$ cm$^{-2}$. $^{(3)}$ Photon index. $^{\star}$Fixed value. $^{(4)}$Temperature of the thermal emitting plasma $^{(5)}$Equivalent width of the narrow  Fe emission line in eV.  $^{(6)}$ Unabsorbed 0.5--2 keV luminosity  in 10$^{42}$ erg s$^{-1}$ (fraction due to the thermal component in parenthesis). $^{(7)}$ Unabsorbed 2--10 keV luminosity  in 10$^{42}$ erg s$^{-1}$. $^{(8)}$2--10 keV observed flux in 10$^{-12}$ erg cm$^{-2}$ s$^{-1}$. $^{\dagger}$The best fit model requires a partially covering absorber, with coverage fraction f$_{c}$=0.41$\pm$0.07. $^{\star}$ This value has been fixed in the spectral fit}.  }
\label{fit}
\end{table*}

\section*{Acknowledgements}
We thank the anonymous referee for his/her helpful comments, which improved the manuscript.
All coauthors, members of the MAGNA project (http://www.issibern.ch/teams/agnactivity/Home.html), gratefully acknowledge support of the International Space Science Institute (ISSI) in Bern, Switzerland.
T.B. acknowledges the support from the Alfred P. Sloan Foundation under Grant No. BR2013-016 and the National Science Foundation under Grant No. NSF AST-1211677.
%

\bsp

\label{lastpage}

\end{document}